\documentstyle[osa,manuscript]{revtex}  
\begin{document}                
\title{THE NEWELL-WHITEHEAD-SEGEL EQUATION FOR TRAVELING WAVES}
\author{Boris A. Malomed}
\address{School of Mathematical Sciences,
Tel Aviv University, Tel Aviv 69978, Israel}
\maketitle
\begin{abstract}                

An equation to describe nearly 1D traveling-waves patterns is
put forward in the form of a dispersive generalization of the 
Newell-Whitehead-Segel equation. Transverse stability of plane waves is 
shown to be drastically altered by the dispersion. A necessary 
{\it transverse Benjamin-Feir} stability condition is obtained.
If it is met, a quarter of the plane-wave existence band is
unstable, while three quarters are transversely stable. Next, linear defects 
in the form of grain boundaries (GB's) are
studied. An effective Burgers equation is derived, in the framework
of which a GB is tantamount to a shock wave. Asymmetric GB's are moving 
at a constant velocity. 
\end{abstract}

\bigskip
PACS numbers (``American"): 47.27.Te; 47.52.+j; 03.40.Kf

\newpage
\section{Introduction}               

The Newell-Whitehead-Segel (NSW) equation is a well-known universal equation
to govern evolution of nearly one-dimensional (1D) nonlinear patterns produced 
by a finite-wavelength instability in isotropic two-dimensional media, a 
classical example being the Rayleigh-B\'{e}nard convection \cite{NWS}. This 
equation was applied to a number of problems, e.g., description of point 
defects \cite{defect} and of linear ones in the form of the so-called grain 
boundaries (GB's) \cite{GB}. 

Thus far, the NWS equation was not extended to the case of traveling waves, 
i.e., to patterns produced by an oscillatory instability, a well-known example 
being the traveling-wave convection in a binary fluids (see, e.g., \cite{TW})
(though, another extension was developed \cite{new1,new2}, aimed at casting
the the usual NWS equation into a rotationally covariant form).
An objective of this work is to put forward a {\it dispersive} NWS equation
for the description of nearly 1D traveling-wave patterns, and apply it to 
relatively
simple problems, viz., transverse stability of plane waves, and GB's. It will be 
demonstrated that the dispersive terms strongly alter the results in comparison
with the usual NWS equation. In particular, all
the plane waves turn out to be transversely unstable if a special condition of 
the Benjamin-Feir type is not satisfied; in the opposite case, the transverse
perturbations destabilize exactly a {\it quarter} of the plane waves' existence 
band. Recall that, for the usual NWS equation, a well-known result is that a 
{\it half}
of the band is transversely unstable against the so-called zigzag perturbations
\cite{NWS}. 

It is relevant to note that experiments with the traveling-wave
convection were usually conducted
in effectively 1D geometries (narrow channels) \cite{TW}. Recently, detailed 
experimental results were obtained for 2D patterns in a large-aspect-ratio 
traveling-wave
convection cell \cite{CA} (a theoretical analysis of the global patterns 
containing GB's was also presented in \cite{CA} on the basis of the 
free-energy functional for the real 2D Swift-Hohenberq equation). These results 
demonstrate stable existence of,
generally speaking, moving GB's separating domains with different orientations
of the traveling waves. This experimental observation was an incentive to 
develop the analysis presented below. 

The NWS equation can only apply to description of the special case when
the waves have a small ``refraction angle" at the GB. In the general case, the 
description 
of a quasi-stationary GB (the one that may be moving, but at a constant
velocity) is based upon coupled Ginzburg-Landau (GL) equations \cite{me}. 
Although the case comprised by the NWS equation is more restricted, a 
more far-reaching analysis will be developed for this case. An effective
Burgers equation governing dynamics of GB's between the traveling waves will 
be derived, which allows one to describe not only a steadily moving GB, but
also certain transient processes and interactions between parallel GB's.

\section{The Dispersive Newell-Whitehead-Segel Equation}

As a starting point, one can take a general two-dimensional equation for a
complex order parameter $u$ in an isotropic medium (cf., e.g., the derivation
of the rotationally invariant extension of the usual NWS equation in 
\cite{new2}):
\begin{equation}
u_t=\epsilon u-\left(\nabla^2+1\right)^2u+if(\nabla^2)u-c_1|u|^2u
-c_2|\nabla u|^2u-c_3(\nabla u)^2u^{*}.
\end{equation}
Here, $\epsilon$ is a small overcriticality parameter, $0<\epsilon\ll 1$,
the operator $(\nabla^2+1)^2$ assumes, as usual, that the instability onset
takes place (at $\epsilon=0$) isotropically at the wave vectors ${\bf k}^2=1$, 
the term $f(\nabla^2)u$ with an arbitrary real-valued operator function 
$f(\nabla^2)$ accounts for the linear
dispersion in a general form, and, under the standard assumption of the weak
nonlinearity, the cubic terms with arbitrary complex coefficients $c_1$, $c_2$,
and $c_3$ account for nonlinear saturation of the 
instability and for nonlinear dispersion. The latter three terms
exhaust all essentially independent local cubic terms in an isotropic system 
near the instability threshold \cite{ZS}. 

Following the derivation of the
classical NWS equation \cite{NWS}, one considers solutions of the form 
$u(x,y,t)=U(x,y,t)\exp(ix+if(-1)t)$, where the envelope $U$ is ssumed to be a 
slowly varying function in comparison with the rapidly
oscillating carrier wave $\exp(ix+if(-1)t)$. Then, a straightforward asymptotic
expansion leads to the following equation for the envelope function:
$$U_t=\epsilon U-\left(2i\partial_x+\partial_y^2\right)^2U+i\left[f^{\prime}(-1)
\left(2i\partial_x+\partial_y^2\right)+\left(f^{\prime}(-1)-2f^{\prime\prime}
(-1)\right)\partial_x^2\right]U$$
\begin{equation}
-\left(c_1+c_2-c_3\right)|U|^2U.
\end{equation}
The convective term (the one with the first derivative) can be eliminated by a
change of the reference frame, $x\rightarrow x-2f^{\prime}(-1)t$. Next, the
small overcriticality $\epsilon$ can be {\it partially} scaled out by means of 
the transformation $x\rightarrow x/\sqrt{\epsilon},\,y\rightarrow y/
\sqrt{\epsilon}$. Here, one notices a drastic difference from the scaling
transformation inherent in derivation of the classical NSW equation: the
presence of the linear dispersion dictates to choose the same 
power of $\epsilon$ in rescaling the longitudinal and transverse coordinates
$x$ and $y$, while, in the usual case, the scaling factor for $y$ would be
$\epsilon^{-1/4}$ instead of $\epsilon^{-1/2}$. 

Additionally rescaling the variables in an obvious way and properly 
redefining the coefficients, one arrives at the following final form of the
dispersive NWS equation:
\begin{equation}
U_t=U-\left(i\partial_x+\sqrt{\epsilon}\,\partial_y^2\right)^2U-(1+i\alpha)
|U|^2U+i\beta U_{xx}+i\gamma U_{yy},\label{NWS}
\end{equation}
where $\alpha$, $\beta$, and $\gamma$ are, respectively, real effective 
coefficients
of the nonlinear, longitudinal linear, and transverse linear dispersions. It
is important that, generically, these coefficients are $\sim 1$, i.e., they do 
not depend 
upon $\epsilon$, on the contrary to the coefficient in front of $\partial_y^2$
in the linear operator in the second term on the r.h.s.
of Eq. (\ref{NWS}). Obviously, the dispersive NWS equation, on the contrary to
the usual one (including the rotationally invariant generalizations
\cite{new1,new2}) does not admit a gradient representation.

It should be stressed that Eq. (\ref{NWS}) is but a simplest model for the
propagation of nearly 1D waves in a nonlinear medium. Indeed, this equation
takes into account only unidirectional waves, ignoring the waves which may
travel in the opposite direction. One may expect that the waves traveling in
both directions will be described by two coupled NWS equations, similarly to the
coupled GL equations \cite{Cross}. The corresponding
analysis will be, of course, more complicated. Another simplification is the
tacit omission of a possible additional real equation for a slowly 
relaxing mode, which represents concentration field in the binary-fluid
convection \cite{Riecke} and occurs in other contexts \cite{other}. It is
known that interaction with this additional real mode can essentially alter
properties of the solutions, especially their stability. Consideration
of a generalized system, coupling the NWS equation to the slow mode, is 
defered to another work. 

\section{Stability of Plane Waves}

As it is implied by the derivation of the NWS equation, an unperturbed plane
wave can be chosen with the zero $y$-component of its wave vector:
\begin{equation}
U_0(x,t)=a_0\,e^{ikx-i\omega t},\;a_0^2=1-k^2,\;\omega=\alpha a_0^2+\beta k^2. 
\label{plane}
\end{equation}
The plane waves
(\ref{plane}) exist inside the band $k^2<1$. It is of obvious interest to
find which solutions inside this band are stable against small perturbations.

If one confines 
the stability analysis to pure longitudinal ($y$-independent) perturbations,
Eq. (\ref{NWS}) is nothing else but the familiar complex GL
equation, in which stability criteria for the plane waves are well known
\cite{GL}. Therefore, it makes sense to concentrate here on consideration of
pure transverse ($x$-independent) perturbations. This amounts to substituting
into Eq. (\ref{NWS}) 
\begin{equation}
U(x,y,t)=e^{ikx-i\beta k^2t}V(y,t),\label{subst}
\end{equation}
which yields the equation
\begin{equation}
U_t=U-\left(k-\sqrt{\epsilon}\,\partial_y^2\right)^2U-(1+i\alpha)|U|^2U
+i\gamma U_{yy}.\label{CSH}
\end{equation}
Of course, the consideration of the weakly overcritical case, $\epsilon\ll 1$,
implies $k^2\ll 1$.

Notice that, at $k<0$, Eq. (\ref{CSH}) is the so-called complex Swift-Hohenberg 
equation, which was first introduced in \cite{ZS}, and which has recently 
attracted a lot of attention as, e.g., a model dynamical equation for laser 
cavities \cite{Jerry}.

The standard approach to
the stability problem is based on presenting a general solution to Eq.
(\ref{CSH}) in the form
$U=a\,\exp(i\phi)$, with the separated amplitude and phase, in terms
of which
the plane-wave solution (\ref{plane}) takes the form $a=a_0$, $\phi=-\omega t$. 
Next, one replaces Eq. (\ref{CSH}) by a system of coupled real equations for 
these 
variables and linearizes them with respect to small perturbations $\delta a$ and
$\delta\phi$ of the amplitude and phase. Finally, one is seeking for 
perturbation eigenmodes is the formally complex form,
\begin{equation}
\delta a =a_1e^{\sigma t+ipy},\,\delta\phi=\phi_1e^{\sigma t+iqy},\label{pert}
\end{equation}
where $a_1$ and $\phi_1$ are infinitesimal amplitudes of the perturbation,
$q$ is an arbitrary real wave number (independent of $k$ in Eq. (\ref{plane})), 
and $\sigma(p)$ is the instability growth rate sought for. 

After a straightforward algebra, one can cast the resolvability condition for
the linearized perturbation equations into the form
$$\sigma^2+2\sigma\left(a_0^2+2k\sqrt{\epsilon}\,q^2+\epsilon q^4\right)$$
\begin{equation}
+q^2\left[\epsilon^2q^6+4k\epsilon^{3/2}q^4+\left(2a_0^2\epsilon+4k^2\epsilon+
\gamma^2\right)q^2+4k\sqrt{\epsilon}\,a_0^2+2\alpha\gamma a_0^2\right]=0.
\label{sigma}
\end{equation}
Because Eq. (\ref{sigma}) is a quadratic equation, the stability condition,
${\rm Re}\,\sigma(q)\leq 0$, is tantamount to demanding that the coefficient 
in front of the linear term and the free term must
be non-negative at all real $q$. In the absence of the dispersion, the 
positiveness of the free term immediately leads to the well-known 
condition $k\geq 0$, whilst the solutions with $k<0$ are unstable against the 
so-called zigzag perturbations \cite{NWS}. In the present case, however, the
dispersion drastically alters this result. Indeed,
taking into regard that $\epsilon$ is a small parameter, while
the coefficients $\alpha$ and $\gamma$ are not small, one obtains the 
following stability condition providing for the positive definiteness of the 
free term in Eq. (\ref{sigma}):
\begin{equation}
\alpha\gamma >0,\label{BF}
\end{equation}
which is, as a matter of fact,  a transverse version of the Benjamin-Fair (BF)
stability condition well known for the longitudinal perturbations in the 
GL equation \cite{GL}. 

Proceeding to the condition for the positive definiteness of the 
coefficient in front of the linear term in Eq. (\ref{sigma}), one sees that
this is automatically satisfied if $k\geq0$. That is why this condition did not
produce any additional restriction in the case of the usual NWS
equation. In the present case, both positve and negative $k$ may be stable
if the inequality (\ref{BF}) holds (and both are unstable if it does not hold).
A simple analysis shows that, at negative $k$, the coefficient in question 
remains positive at all values of $q$ if $k^2<a_0^2$, or, finally, with regard
to Eq. (\ref{plane}), if
\begin{equation}
k\geq -1/\sqrt{2}.\label{quarter}
\end{equation} 

It is easy to check that the dispersion equation (\ref{sigma}) does not produce
any additional stability condition but (\ref{BF}) and (\ref{quarter}).
Thus, the eventual result of the stability analysis for the transverse
perturbations is: all the plane waves are unstable if the transverse BF
condition (\ref{BF}) does not hold; in the opposite case,
the transversely unstable subband inside the existence band, $-1<k<+1$,
is (according to Eq. (\ref{quarter}) $-1<k<-1/\sqrt{2}$. Thus, while in
the classical NWS equation the transverse perturbations render a half of the 
existence band, $k<0$, unstable \cite{NWS}, in the dispersive case they 
destabilize exactly its {\it quarter} (in terms of $k^2$), provided that the BF
condition (\ref{BF}) is satisfied. 

\section{The Grain Boundaries}

The consideration of GB patterns in two-dimensional nonlinear dissipative
media has attracted a lot of attention starting from the works 
\cite{first}. An analytical description of GB's between domains (``grains") 
occupied by
plane waves with different orientations is usually possible when the 
problem can be reduced to an effectively 1D one \cite{GB,me}. In the case 
of Eq. (\ref{NWS}), there are two natural ways to arrive at a 
1D problem: to substitute either $U(x,y,t)=e^{iqy}V(x,t)$ or Eq. (5),
with an arbitrary constant wave number $q$ or $k$. The former substitution, 
obviously, brings Eq. (\ref{NWS}) into the form of the traditional complex
GL equation. In terms of this equation, the GB will correspond to a
shock-wave solution, which is produced by a collision of traveling waves with 
different wave numbers \cite{me2,GB,me}. This solution
can be obtained in an approximate analytical form by means of the 
phase-diffusion equation \cite{me2}. Coming
back to the interpretation of this solution as a GB in 2D patterns, one will 
have the GB nearly parallel to the ``crests" of the colliding waves (recall 
that the transition from
Eq. (1) to Eq. (\ref{NWS}) implies separation of the slowly varying envelope
and rapidly varying carrier wave whose wave vector is parallel to the $x$
axis). The linear defect of this type should be, actually, described as an 
array of dislocations \cite{GB} rather than a smooth GB.

The substitution (\ref{subst}), which transforms Eq. (\ref{NWS}) into Eq. 
(\ref{CSH}), leads to essentially new results for smooth GB's. In
this case, one should expect a GB exactly or nearly perpendicular to 
the wave crests, with a small ``refraction angle" at the GB, see Fig. 5 in
\cite{GB}. To describe the GB of this type in an analytical form, it is
natural to consider, first of all, the lowest-order approximation, omitting 
the small terms (which vanish along with $\epsilon$) in Eq. (\ref{CSH}).
This approximation leads to a simple but, nevertheless, nontrivial equation,
viz., a special form of the complex GL equation without the diffusion term:
\begin{equation}
U_t=(1-k^2)U-(1+i\alpha)|U|^2U+i\gamma U_{yy}.\label{GL}
\end{equation}

In terms of Eq. (\ref{GL}), the GB again corresponds to a shock wave. As well as
in the previous section, one should substitute $U=a\,\exp(i\phi)$, to transform
Eq. (\ref{GL}) into the coupled real equations,
\begin{eqnarray}
a_t=(1-k^2)a-a^3-2\gamma a_yq-\gamma aq_{y},\label{eqa}\\
a\phi_t=-\alpha a^3+\gamma a_{yy} -\gamma aq^2,\label{phi}
\end{eqnarray}
where $q\equiv\phi_y$. 
Following the usual assumption on which the phase-diffusion approximation
(called the geometric optics approximation in \cite{me2}) is based, i.e., that
the local amplitude $a$ and the local wave number $q$ are varying
at a large spatial scale, one may neglect, in the zeroth-order approximation, 
the spatial derivatives on the right-hand sides of Eqs. (\ref{eqa}) and 
(\ref{phi}). In this approximation, Eq. (\ref{eqa}) yields $a=\sqrt{1-k^2}$.
A straightforward consideration shows that the
next-order term in Eq. (\ref{eqa}) is $-\gamma aq_y$. Taking it into account
as a small perturbation, one obtains the corrected expression for the 
amplitude:
\begin{equation}
a=\sqrt{1-k^2}-\frac{\gamma}{2}\left(1-k^2\right)^{-1/2}q_y.\label{a}
\end{equation}
This should be inserted into Eq. (\ref{phi}). Prior to this, it is convenient
to transform Eq. (\ref{phi}), omitting the last term which is of a higher order
of smallness, dividing it by $a$, and differentiating the resultant equation
in $y$, so that to eliminate the phase by means of the identity $\phi_{ty}
\equiv q_t$. This procedure produces the equation
\begin{equation}
q_t=-2\alpha aa_{y}-2\gamma qq_y.\label{q}
\end{equation}
At last, substitution of Eq. (\ref{a}) into Eq. (\ref{q}) yields, in the lowest 
nontrivial approximation, exactly the classical Burgers equation \cite{With}
\begin{equation}
q_t=(\alpha\gamma)q_{yy}-2\gamma qq_y.\label{Bur}
\end{equation}
Notice that this equation is well-posed if the above stability condition
(\ref{BF}) holds. It is noteworthy that the combination $(1-k^2)$ does not show 
up in Eq. (\ref{Bur}).

Before proceeding to consideration of its relevant solutions, it is also
interesting to consider an extended equation produced by adding higher-order
terms to Eq. (\ref{Bur}). First, one can restore the small terms $\sim
\sqrt{\epsilon}$ and $\sim\epsilon$, that were omitted when deriving Eq. 
(\ref{GL}) from Eq. (\ref{CSH}). Because these terms are already small, it 
will be sufficient to calculate them at the lowest order in the sense of the
phase-diffusion equation. Next, the second term on the right-hand side of
Eq. (\ref{eqa}) produces the next-order phase-diffusion correction to Eq. 
(\ref{a}). Collecting all the corrections, Eq. (14) becomes
$$a=\sqrt{1-k^2}-\frac{\gamma}{2}\left(1-k^2\right)^{-1/2}q_y-\left(1-k^2
\right)^{-1/2}\sqrt{\epsilon}\,q^2\left(k+\frac{1}{2}\sqrt{\epsilon}\,q^2
\right)$$
\begin{equation}
-\frac{\gamma^2}{2} \left(1-k^2\right)^{-3/2}qq_{yy}.\label{acorr}
\end{equation}
Finally, insertion of this equation into Eq. (\ref{q}) leads to the following
extended Burgers equation:
$$q_t=(\alpha\gamma)q_{yy}-2\left(\gamma-2\alpha\sqrt{\epsilon}\,k\right)qq_y$$
\begin{equation}
+4\epsilon\alpha q^3q_y-\gamma^2\left(1-k^2\right)^{-1}\left[\alpha
\left(qq_{yy}\right)_y+\frac{1}{2}q_yq_{yy}+q_{yyy}\right].\label{Burcorr}
\end{equation}

The Burgers equation (\ref{Bur}) can be exactly solved by means of the Cole-Hopf
transformation \cite{With}.
In particular, its simplest nontrivial solution describes a steadily moving
shock wave:
\begin{equation}
q(y,t)=q_1+\frac{q_2-q_1}{1+\exp\left(\alpha^{-1}(q_2-q_1)(y-ct)\right)},
\label{shock}
\end{equation}
where $q_1$ and $q_2$ are the asymptotic values of the local wave number
at $y=\pm\infty$ (for the definiteness, it is assumed $q_2>q_1$), and the shock 
wave's velocity is
\begin{equation}
c=\gamma(q_1+q_2).\label{c}
\end{equation}

To interpret this solution in terms of the underlying 2D traveling-waves
pattern,
one should recollect the expression for the full local wave vector implied
in the derivation of Eqs. (\ref{CSH}) and (\ref{GL}). According to Eq.
(\ref{subst}), it is $(1+k,q)$, with sufficiently small $k$ and $q$. This
means that the ``crests" of the waves which are forming the pattern are nearly 
parallel 
to the $y$ axis, while the GB itself is strictly perpendicular to this axis.
The corresponding picture is essentially tantamount to Fig. 5 in \cite{GB}.
The asymptotic angles between the crests and the axis $y$ (the normal to the GB)
at $y\rightarrow\pm\infty$ are, obviously, $\approx q_{1,2}$. If $q_1=-q_2$,
the pattern is symmetric, the GB being its symmetry axis, and, according to Eq.
(\ref{c}), its velocity is zero. An asymmetric GB is possible as well, moving
at a nonzero velocity given by Eq. (\ref{c}). The same general inference was
formulated in \cite{me}, where GB's between traveling waves with {\it nonsmall} 
``refraction angles" were considered within the framework of coupled complex GL 
equations. It is also interesting to compare this situation with that described 
by the usual NWS equatiton. For that case, it was demonstrated in \cite{GB}
(section III.E) that a stationary GB was possible only in the strictly 
symmetric case ($q_1=q_2$), and only under a special condition that the 
asymptotic
wave numbers lay exactly at the boundary of the zigzag instability, i.e., at
$k=-q_{1,2}^2$. As it was demonstrated in this section, a drastic difference
of the GB between the traveling waves is that it exists, within the framework 
of applicability of Eq. (\ref{NWS}), as a generic solution. This qualitative
result is also in agreement with what was obtained in \cite{me} for the GB's
with nonsmall refraction angles. It is relevant to mention that, in models of
nonoscillatory media, a stationary GB with a nonsmall refraction angle may be 
asymmetric, but the corresponding asymptotic wave numbers must be exactly equal 
to one \cite{GB}.

Coming back to the effective Burgers equation (\ref{Bur}), one can notice that
the integrability of this equation allows to obtain not only the 
shock-wave solution (\ref{shock}), but also explicit solutions describing 
transients and
interactions between the shock waves, or in terms of Eq. (\ref{NWS}), 
interactions of the parallel GB's. In particular, solutions describing 
formation of a GB out of various initial perturbations, as well as a
collision of two shock waves and their merger into a single one, are available
\cite{With}. Although a more detailed comparison with experimental observations
of the traveling waves in the large-aspect-ratio binary-fluid convection cells
\cite{CA} is necessary, it appears that formation of a GB and merger of GB's as
a result of their interaction can be observed.

It may be interesting to analyze influence of the small additional 
terms in the extended Burgers equation (\ref{Burcorr}) on properties of
the shock waves, but this issue is left beyond the framework of this work.
Lastly, it is relevant to mention that the special GL equation (\ref{GL}) has
an altogether different physical application in terms of nonlinear fiber optics
{\cite{HK}).
If $t$ and $y$ are interpreted, respectively, as the propagation distance and
the so-called reduced time in a fiber, this is a propagation equation for the
light in the presence of a broadband gain and linear and nonlinear losses
\cite{Jena}. The condition (\ref{BF}) is then equivalent to the condition that 
the fiber's dispersion is normal. In this case, cw (continuous-wave) states with
different frequencies $q$ are stable, and the shock wave solutions, e.g.,
(\ref{shock}),
describing collisions between the cw's, are of a certain physical interest.

\section{ACKNOWLEDGEMENTS}

I appreciate stimulating discussions with C.M. Surko and A. La Porta. I 
especially appreciate a possibility to see their experimental results prior
to the publication. I am also indebted to J.V. Moloney and A.A. Nepomyashchy 
for useful discussions of related problems.

\end{document}